# Control and Synchronization of Chaotic Fractional-Order Coullet System via Active Controller

M. Shahiri T.*, A. Ranjbar N.*,
R. Ghaderi*, S. H. Hosseinnia*, S. Momani**

* *Noushirvani University of Technology, Faculty of Electrical and Computer Engineering, P.O. Box 47135-484, Babol, Iran,*
*(a.ranjbar@nit.ac.ir) ,(h.hoseinnia@stu.nit.ac.ir)*
** *Department of Mathematics, Mutah University, P.O. Box: 7, Al-Karak, Jordan*

___________________________________________________________

**Abstract:** In this paper, fractional order Coullet system is studied. An active control technique is applied to control this chaotic system. This type of controller is also applied to synchronize chaotic fractional-order systems in master–slave structure. The synchronization procedure is shown via simulation. The boundary of stability is obtained by both of theoretical analysis and simulation result. The numerical simulations show the effectiveness of the proposed controller.

___________________________________________________________

## 1. INTRODUCTION

Fractional calculus has 300-year history. However, applications of fractional calculus in physics and engineering have just begun (Hilfer, 2001). Many systems are known to display fractional order dynamics, such as viscoelastic systems (Bagley and Calico, 1991), dielectric polarization, and electromagnetic waves. In the recent years, emergence of effective methods in differentiation and integration of non-integer order equations makes fractional-order systems more and more attractive for the systems control community. The TID controller (Lune, 1994.), the fractional PID controller (Podlubny, 1999), the CRONE controllers (Oustaloup, 1999 ,1996, 1995) and the fractional lead-lag compensator (Raynaud and Zerga, 2000; Monje and Lille, 2004.) and Simple Fractional Controller (Tavazoei and Haeri, 2008), are some of well-known fractional-order controllers .

More recently, there has been a new trend to investigate the control and dynamics of fractional order chaotic systems (Ahmad, 2003a, b, c ; Li, 2004, 2003). It has been shown (Ahmad, 2003b) that nonlinear chaotic systems may keep their chaotic behavior when their models become fractional. In (Ahmad, 2003a), chaos control was successfully investigated for fractional chaotic systems, where controllers have been designed using a ''backstepping'' technique. It was demonstrated that nonlinear controllers, which is designed to stabilize the integral chaotic model, might still stabilize the fractional order model (Ahmad, 2004). It is simultaneously shown that chaos exists in the fractional order Chen system when the order is less than 3. A Linearizing feedback technique has successfully been applied on this chaotic system. In (Li et. al, 2003), synchronization of fractional order chaotic systems has been studied. Utilizing an approximation approach of fractional operator, the system behaves chaotic, when the order is less than number of state i.e. 3. Chaos control and synchronization have widely been studied for an integral order system (Ge and et. al., 2005a, b, 2004a, b, c, d). The work will be generalized for a fractional order Coullet system. Synchronization of Coullet system is reported in (Jian-Bing et. al., 2008) using backstepping method. There is still lack of report to control and synchronize a fractional-order Coullet system. In this paper, the aim is to control and synchronize two chaotic fractional-order Coullet systems, using an Active Control.

This paper is organized as follows:
Coullet system will be described in section 2. Active controller will be presented in section 3. This controller is applied to synchronize two identical fractional-order Coullet systems in section 4. Ultimately, the work will be concluded at section 5.

## 2. SYSTEM DISCRIPTION

Some of nonlinear systems represent a chaotic behaviour. These systems are very sensitive to initial conditions. This means two identical distinct systems but with a minor deviation in their initial condition may result completely different. This means having known bounded initial conditions, there is less chance for the dynamics to predict the behaviour. This deterministic treatment is called chaos.

Consider the following Arnéodo–*Coullet* dynamics equation (Arnéodo et. al., 1981)**:**

$$\dddot{x} + a\ddot{x} + b\dot{x} + cx + dx^3 = 0 \quad (1)$$

which exhibits chaotic dynamics for various values of four parameters (Coullet et. al., 1979; Arnéodo et. al., 1981). The system is simulated for the following set {a=0.8, b=-1.1, c= -0.45, d=-1}.

The system in (1) will be written in the state space format as:

$$\begin{cases} \dot{x}_1 = x_2 \\ \dot{x}_2 = x_3 \\ \dot{x}_3 = cx_3 + bx_2 + ax_1 + dx_1^3 \end{cases} \quad (2)$$

A phase portrait of the system in Figure (1) represents a chaotic behaviour.

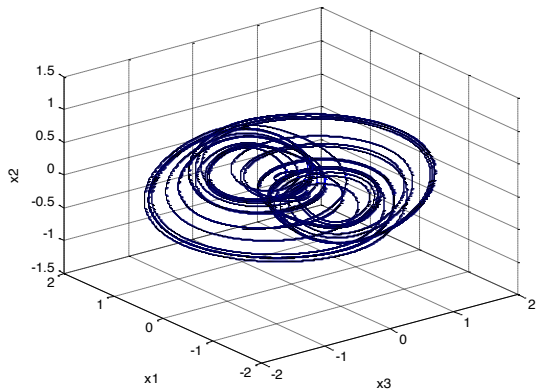

Fig. 1. Phase portrait of Chaotic Coullet system

Although some works have reported on chaotic Coullet system, there is still lack of report on fractional order Coullet system. In this paper, an active controller as a new control system is applied on fractional order chaotic system.

To define the fractional order Coullet chaotic system the equation will be defined as (3), in which $q$ will changed in accordance to fractional order of defined system. It can be shown that for some range of $q$, the fractional order Coullet system is unstable (Tavazoei and Haeri, 2008). A resonance property of fractional order Coullet chaotic system is shown in Figure (3) for different values of $q=0.97, 0.95,$ and $0.9$.

The numerical simulations have carried out using the SIMULINK based on the frequency domain approximation.

$$\begin{cases} D^q x_1 = x_2 \\ D^q x_2 = x_3 \\ D^q x_3 = cx_3 + bx_2 + ax_1 + dx_1^3 + u(t) \end{cases} \quad (3)$$

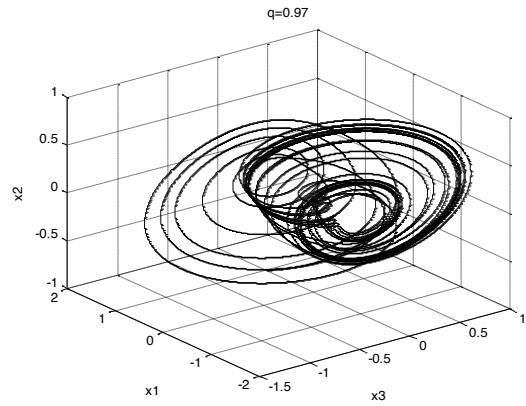

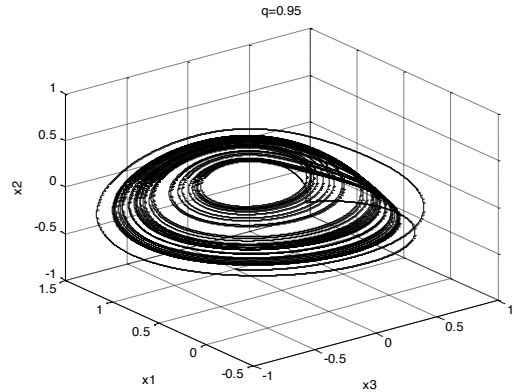

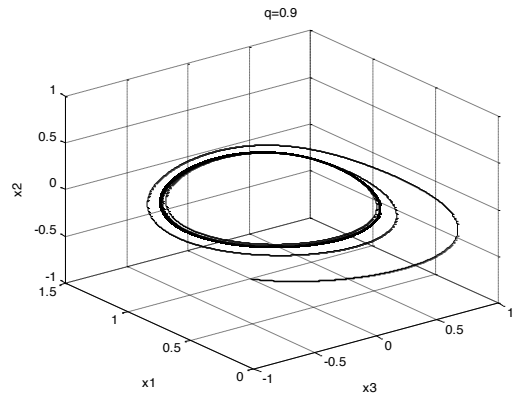

Fig. 2. Phase portrait of chaotic Coullet system for different values of the fraction parameter.

In continuing the system will be stabilized using an active control.

3. ACTIVE CONTROL OF COULLET SYSTEM

In the active control method, the control signals are directly added to the fractional-order system dynamics as:

$$\begin{cases} D^q x_1 = x_2 \\ D^q x_2 = x_3 \\ D^q x_3 = f(x) + u(t) \end{cases} \quad (4)$$

The control signals $u$ constructed from two parts. The first part is considered to eliminate the nonlinear part of equation (4). The second part $v(t)$ acts as external input in (4) which is designed to stabilize the system:

$$u(t) = -f(x) + v(t) \quad (5)$$

where,

$$v = -k_1 x_1 - k_2 x_2 - k_3 x_3 \quad (6)$$

where, $k_i \geq 0$ is chosen according to the designer. The design procedure consists of spotting gains $k_i$ to stabilize dynamics in (4). In the following, relevant matrices and relations are given for Control of Coullet chaotic system. Now, regarding the fractional order Coullet system in equation (4), the control input is determined as:

$$u(t) = -cx_3 - bx_2 - ax_1 - dx_1^3 + v \quad (7)$$

Dynamics in (7) can be represented by a fractional state space dynamics:

$$\begin{cases} D^q x_1 = x_2 \\ D^q x_2 = x_3 \\ D^q x_3 = v \end{cases} \quad (8)$$

A proper adjustment of the gain parameter locates unstable eigenvalue to a stable position. The equation will be rewritten in the following form:

$$D^q X = AX, A = \begin{bmatrix} 0 & 1 & \cdots & 0 \\ 0 & 0 & 1 \cdots & 0 \\ \cdot & \cdot & & \cdot \\ -k_1 & -k_2 & \cdots & -k_n \end{bmatrix} \quad (9)$$

It is evident that eigenvalues of matrix $A$ spots the stability performance. The stability will be shown in the following for fractional system.

### 3.1 Stability analysis for Fractional order systems

A fractional order linear time invariant (LTIFO) system may be defined in the following state-space format:

$$\begin{cases} D^\alpha x = Ax + Bu \\ y = Cx \end{cases} \quad (10)$$

where, $x \in R^n$, $u \in R^r$ and $y \in R^p$ denote states, input and output vectors of the system will be shown by $A \in R^{n \times n}$, $B \in R^{n \times r}$ and $C \in R^{p \times n}$ respectively, and $q$ is the fractional commensurate order. Fractional order differential equations are at least as stable as their integer orders counterparts, because systems with memory are typically more stable than their memory-less alternatives (Ahmed. E. et al, 2007; Hosein nia et. al, 2008; Ranjbar et. al, 2008). It has been shown that the autonomous dynamics $D^q x = Ax$, $x(0) = x_0$ is asymptotically stable if the following condition is met (Matignon and Lille, 1996):

$$|\arg(eig(A))| > q\pi/2 \quad (11)$$

where, $0 < q < 1$ and $eig(A)$ represents the eigenvalues of matrix $A$. In addition, this system is stable if $|\arg(eig(A))| \geq q\pi/2$ and those critical eigenvalues which satisfy $|\arg(eig(A))| = q\pi/2$ have geometric multiplicity of 1. The stability region for $0 < \alpha < 1$ is shown in Figure 3.

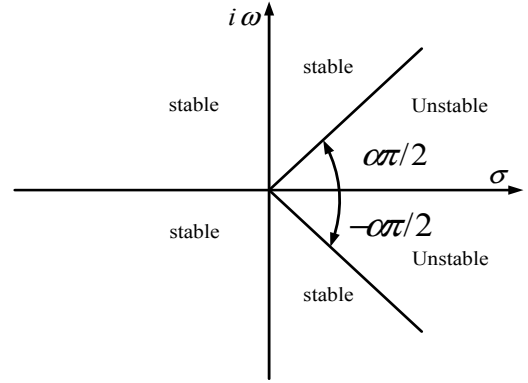

Fig. 3. Stability region of the LTIFO system with fractional order, $0 < \alpha < 1$

Again the gain $K_i$ will be chosen according to equation (9). The stability region will be obtained by equation (11) to stabilize the system.

### 3.2 applying this controller on fractional-order Coullet systems

Let us consider the fractional order Coullet chaotic system regarding the parameters $a=0.8$, $b=-1.1$, $c=-0.45$, $d=-1$, $q=0.97$ and controller $u(t)$:

$$\begin{cases} D^q x_1 = x_2 \\ D^q x_2 = x_3 \\ D^q x_3 = -0.45x_3 - 1.1x_2 + 0.8x_1 - x_1^3 + u(t) \end{cases} \quad (12)$$

From equation (7) the control input of system (12) is determined as:

$$u = +0.45x_3 + 1.1x_2 - 0.8x_1 + x_1^3 + v \quad (13)$$

Initial conditions are chosen as: $x_1(0) = 1$, $x_2(0) = -1$ and $x_3(0) = 0$. Parameters of the controllers are also selected as $k_1 = -10$, $k_2 = -5$ and $k_3 = -4$. The corresponding eigenvalues are obtained as -3.39 and $-0.3026 \pm 1.6894i$, which satisfy the stability condition of $|\arg(eig(A))| > q\pi/2$. This means, the linearizing state feedback and together with proper parameter adjustment stabilizes unstable system.

In this section, numerical simulations have carried out using the SIMULINK based on proper solver. The employed time step size in this simulation is 0.001.

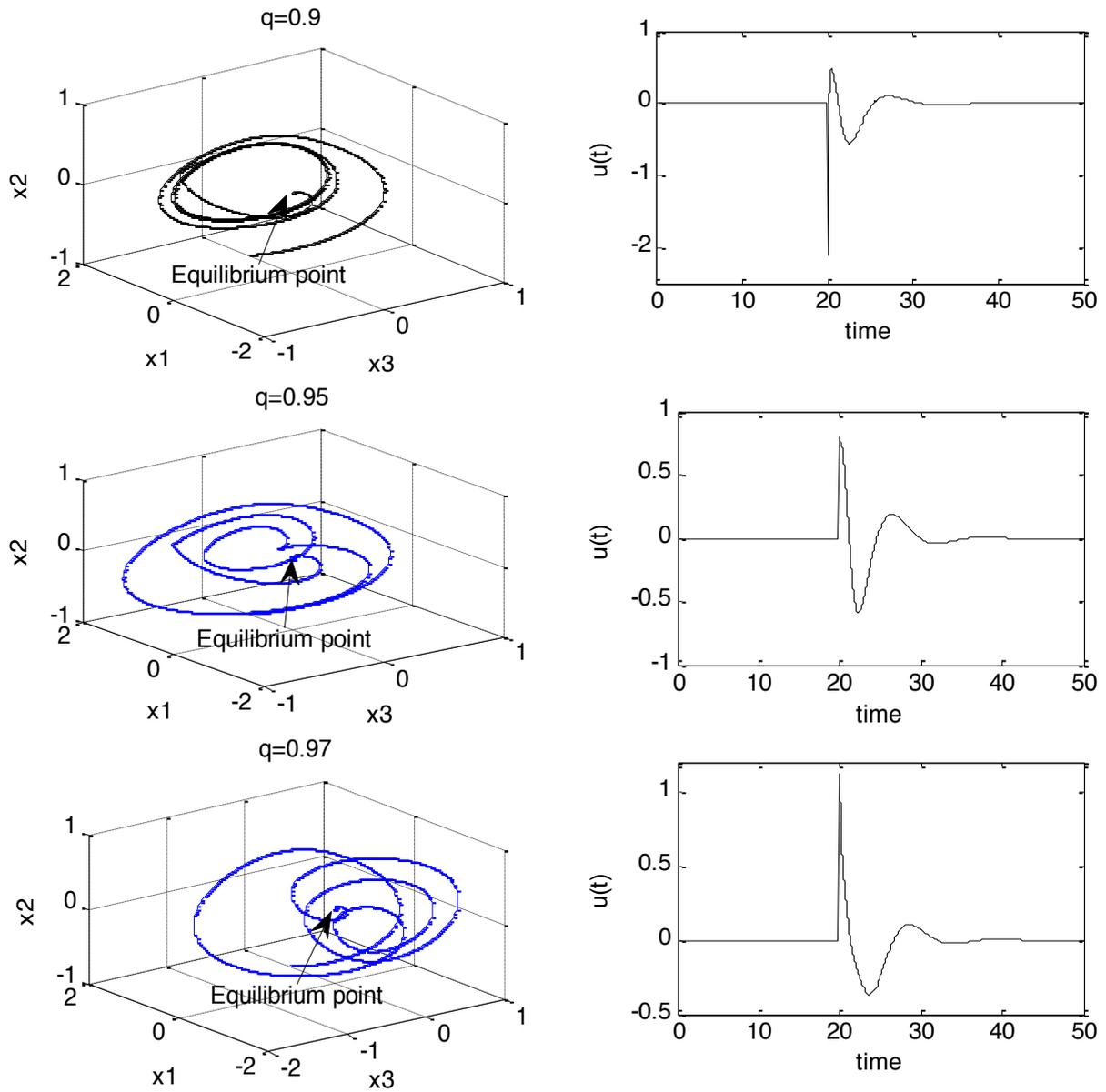

Fig. 4 . Phase portrait of controlled system (left) and the control signal (right) for q = 0.97, 0.95, 0.9

The phase portrait of the controlled system (left) and the appropriate control signal (right) are shown in Figure (4), considering different $q$ ($q=0.97, 0.95, 0.9$). As it can be seen, the system is approached to a stable steady state. Note that the control is activated in t=20 Seconds.

## 4. SYNCHRONIZATION OF FRACTIONAL-ORDER COULLET SYSTEM

Basically, chaos synchronization problem means making two systems oscillate in a same way. Let us call a particular dynamical system *master* and a different dynamical system a *slave*. The goal is to synchronize the slave system with the master using an active controller. In order to achieve this synchronization, a nonlinear control system that obtains signals from the master system and controls the slave system should be designed. Let us consider two fractional order Coullet systems with different initial condition as *master* and *slave* systems respectively:

$$master \begin{cases} D^q x_1 = x_2 \\ D^q x_2 = x_3 \\ D^q x_3 = cx_3 + bx_2 + ax_1 + dx_1^3 \end{cases} \quad (14)$$

$$(x_{10}, x_{20}, x_{30}) = (a_1, b_1, c_1)$$

$$slave \begin{cases} D^q y_1 = y_2 \\ D^q y_2 = y_3 \\ D^q y_3 = cy_3 + by_2 + ay_1 + dy_1^3 + u \end{cases} \quad (15)$$

$$(y_{10}, y_{20}, y_{30}) = (a_2, b_2, c_2)$$

As it has already been mentioned; the control signal should be designed in a way which the slave follows the master. The error is defined as the discrepancy of the relevant states i.e. $e_i = y_i - x_i$ for $i = 1, 2, 3$.

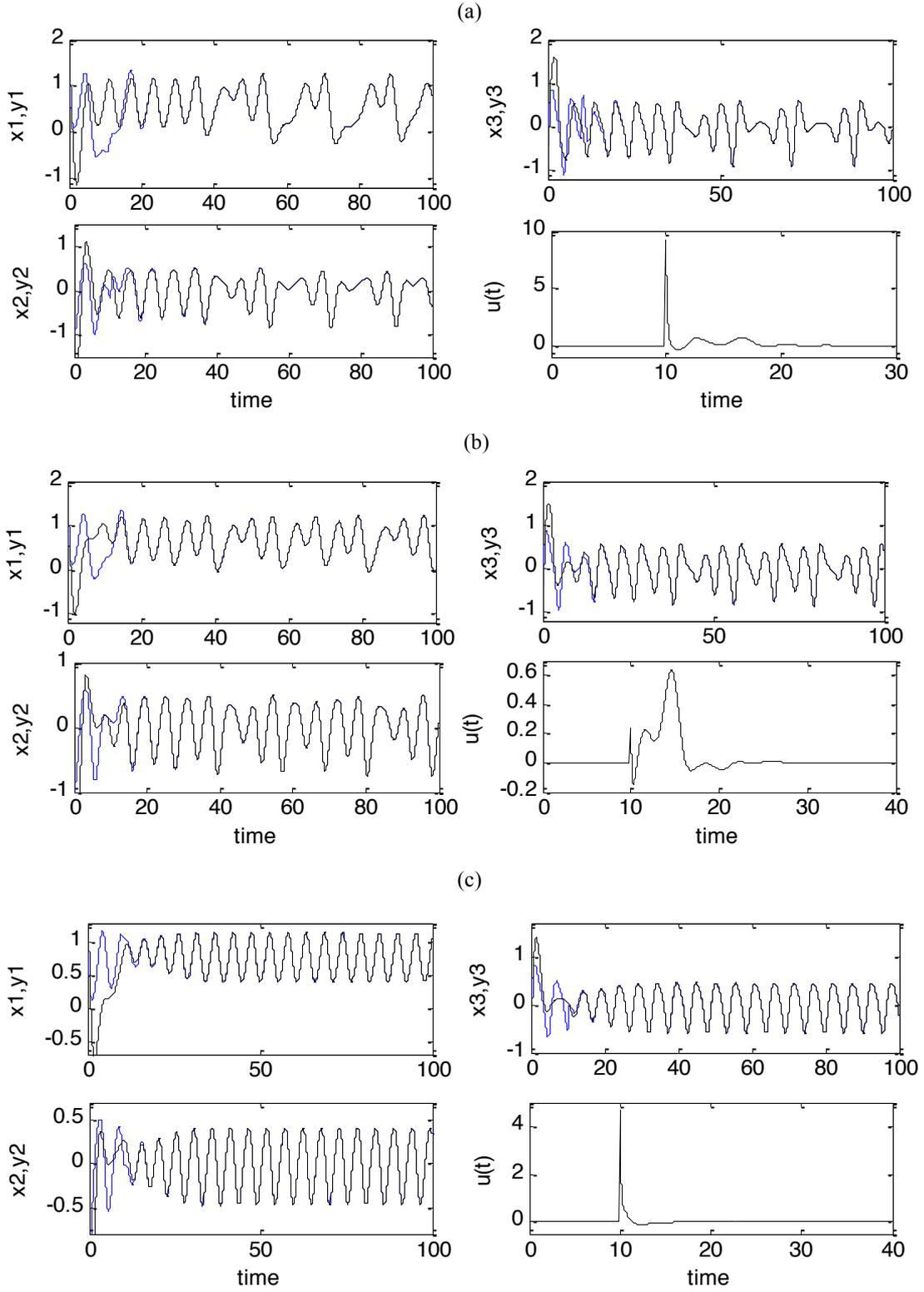

**Figure 5:** The control signal and synchronization of states $x_1, x_2, x_3$ for a: $q=0.97$, b: $q=0.95$, c: $q=0.9$

Deducing the master dynamics from the slave one, approaches us to:

$$\begin{cases} D^q e_1 = e_2 \\ D^q e_2 = e_3 \\ D^q e_3 = ce_3 + be_2 + ae + d(x_1^3 - y_1^3) - u \end{cases} \quad (16)$$

$$e_i = y_i - x_i \quad (17)$$

$$u(t) = -ce_3 - be_2 - ae_1 + d(x_1^3 - y_1^3) + v \quad (18)$$

By choosing the proper value for control parameters ($K_i$), $\lim_{t \to \infty} e = 0$ therefore $x_i = y_i$ and then synchronization between master and slave will be happened.

A linearizing state-input feedback using MATLAB® 7.4 is used to synchronize a fractional Coullet chaotic system. Initial conditions of the master and slave are chosen as $x_1(0) = 1$, $x_2(0) = -1$, $x_3(0) = 0$ and $y_1(0) = 1.5$ $y_2(0) = -1.5$ and $y_3(0) = 0$ respectively.

The control law using parameters as *a=0.8*, *b= -1.1* and *c= -0.45* are obtained considering equation (5) as follows:

$$u(t) = -0.45e_3 - 1.1e_2 + 0.8e_1 - x_1^3 + y_1^3 + v \qquad (19)$$

It should be noted that proper selection of gain parameters $k_i$ causes the error tends to zero, and therefore the slave follows the Master. Simulation result for $k_1 = -4, k_2 = -5$ and $k_3 = -8$ is shown in Figure (5). The control signal and the synchronization of the states $x_1, x_2$ and $x_3$ are shown if Figure (5) for $q = 0.97$, $q = 0.95$ and $q = 0.9$ in (a), (b) and (c) respectively. When the control is activated in $t = 10s$, those two systems are immediately synchronized. The simulation results verify the significance of the proposed controller on the fractional system.

## 4. CONCLUSIONS

In this paper, control and synchronization of chaotic Coullet system with fractional orders is investigated. Active Control method has been chosen to control this chaotic system. It has been shown that by proper selection of the control parameters ($K_i$), the master and slave systems are synchronized. Numerical simulations show the efficiency of the proposed controller in the synchronize task.